\documentclass[a4paper,fleqn,usenatbib]{mnras}

\usepackage[T1]{fontenc}
\usepackage{ae,aecompl}

\usepackage{graphicx}	
\usepackage{amsmath}	
\usepackage{amssymb}	
\usepackage{epsfig}
\usepackage{amssymb}
\usepackage{rotating}

\newcommand{\iue}{{\it IUE}}
\newcommand{\hst}{{\it HST}}
\newcommand{\civ}{C~{\sc iv}}
\newcommand{\siiv}{Si~{\sc iv}}
\newcommand{\mdot}{$\dot{M}$}
\newcommand{\kms}{km s$^{-1}$}
\newcommand{\cpd}{CPD$-41^{\rm o}$}
\newcommand{\bcep}{$\beta$~Cep}
\newcommand{\llep}{$\lambda$~Lep}
\newcommand{\mdotperyr}{M$_\odot$~yr$^{-1}$}

\title[Highly ionized, high velocity gas in NGC 6231]{The highly ionized, 
high velocity gas in NGC 6231}

\author[Derck Massa]{Derck Massa,$^{1}$\thanks{E-mail: dmassa@spacescience.org}
\\
$^{1}$Space Science Institute, 4750 Walnut St., Suite 205, Boulder, Colorado 80301, USA\\
}

\date{Accepted XXX. Received YYY; in original form ZZZ}

\pubyear{2016}

\begin{document}
\label{firstpage}
\pagerange{\pageref{firstpage}--\pageref{lastpage}}
\maketitle

\begin{abstract}
It is well known that clusters of massive stars are influenced by the 
presence of strong winds, that they are sources of diffuse X-rays from 
shocked gas, and that this gas can be vented into the surrounding region or 
the halo through the champagne effect. However, the details of how these 
different environments interact and evolve are far from complete.  This 
paper attributes the broad \civ$\lambda\lambda 1500$ absorption features 
(extending to $-1900$ \kms) that are seen in the spectra of main sequence B 
stars in NGC~6231 to gas in the cluster environment and not the B stars 
themselves.  It is shown that the presence of a WC star, WR~79, in the 
cluster makes this gas detectable because its wind enriches the cluster gas 
with carbon.  Given the available data, it is not clear whether the 
absorbing gas is simply the far wind of WR~79 or a collective cluster wind 
enriched by carbon from the wind of WR~79.  If it is simply due to the 
wind, then this wind must flow, unimpeded for more than 2 pc, suggesting 
that the inner region of the cluster is nearly devoid of obstructing 
material.  If it is actually a collective wind from the cluster, then we 
could be witnessing an important stage of galactic feedback.  In either 
case, the observations provide a unique and significant piece to the 
puzzle of how massive, open clusters evolve.  
\end{abstract}



\section{Introduction}
Wolf Rayet stars have powerful, highly ionized stellar winds, and the 
signature of these winds define their spectra class.  Their winds have been 
imaged on scales of hundreds of AU \citep[e.g.,][]{tut08} and their power is 
evident from the wind blown bubbles they can produce, which can be several 
pc in diameter \citep[e.g.,][]{caz00, naz03}.  In a cluster, these winds may 
carve out large cavities around the star or may interact with the winds of 
other cluster members.  This paper shows that because WR~79 (HD~152270) 
resides in the young, compact open cluster NGC~6231, it is possible to use 
the main sequence B stars (BVs thereafter) in the cluster to probe the 
influence of its carbon rich, massive wind at distances more than a parsec 
from the star.  This wind is identified as the source of strong, high speed 
\civ $\lambda \lambda 1550$ absorption seen in \iue\ spectra of BVs in the 
core of NGC~6231.  The following sections provide: an overview of NGC~6231 
and WR~79, including why a previous explanation for the high speed 
absorption is no longer tenable (\S~\ref{sec:6231}); an explanation of why 
the wind of WR~79 can be observed in NGC~6231 (\S~\ref{sec:ew}); a 
derivation of the absorption profiles expected for BVs embedded in a wind 
(\S~\ref{sec:model}); fits of the model profiles to the 2 BVs in NGC~6231 
that have high dispersion \iue\ spectra (\S~\ref{sec:fits}), and; a 
discussion of the results (\S~\ref{sec:dis}).

\section{NGC 6231}\label{sec:6231}
NGC 6231 is a compact open cluster at the core of Sco OB1, at a distance of 
1.64 kpc \citep{bal95}.  The core of the cluster contains numerous BVs, 10 
O stars, a B supergiant and the WC+O star WR~79.  Its stellar content has 
been well studied \citep[e.g.,][]{sch69, bau99}, and it is a site of ongoing 
star formation \citep{san07}.  Polarimetric observations by \citet{fei03} 
suggest there has been a supernova explosion in the cluster, but they noted 
that their observations are also consistent with an expanding bubble of the 
sort attributed to the WC stars WR~101 and WR~113 by \citet{cap02}.  Red 
DSS2 images of the region show that NGC~6231 is within a large, $\sim 
1.7^\circ \times 3.5^\circ$ ($\sim 50$ pc $\times 100$ pc), cavity centered 
near $\zeta^1$~Sco, and give the impression that the cluster might be 
venting gas from its H~II region.  However, such gas has not been 
detected.  WR~79 is one of the most luminous objects in the cluster.  It is 
a colliding wind binary system whose wind was modeled by \citet{hil00}, who 
showed that the WC component dominates the overall wind flow.  

\begin{figure}
\begin{center}
\includegraphics[width=1.0\linewidth]{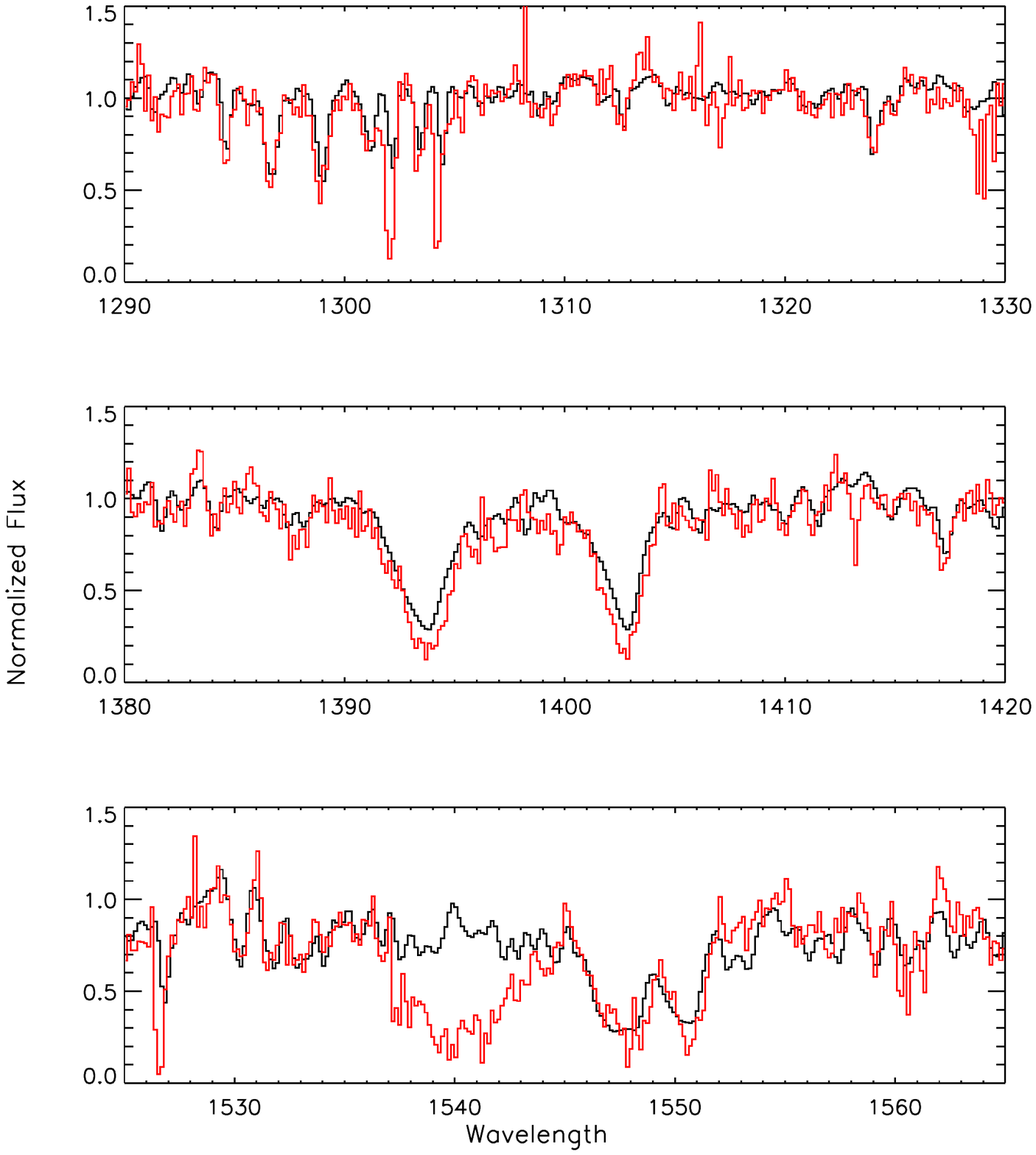}\vfill
\vspace{-1.0in}\includegraphics[width=1.0\linewidth]{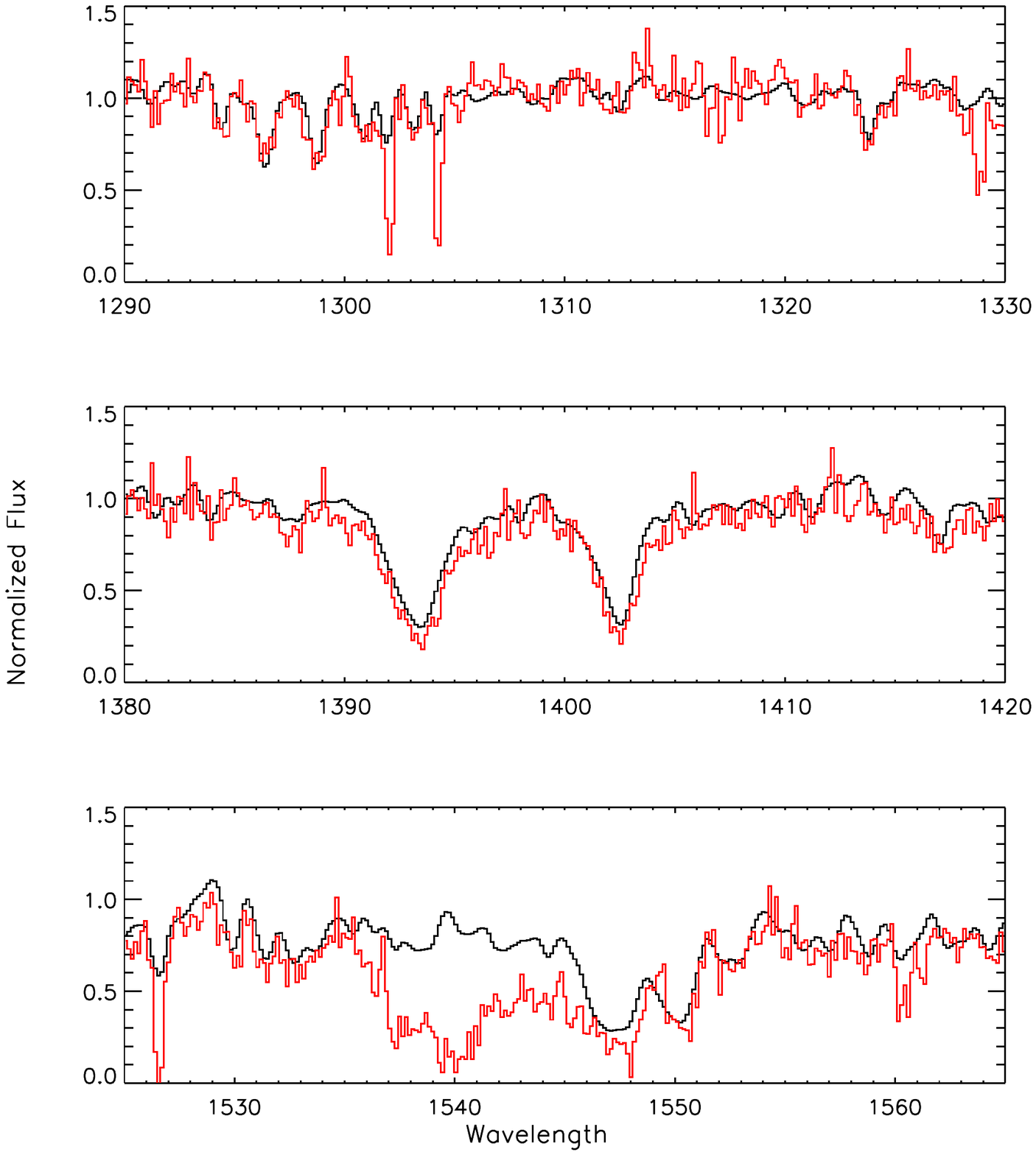}
\vspace{-1.in}\caption{Comparisons of the NGC~6231 BV spectra (red) with 
that of the normal B0.5 IV star \llep\ (black).  Top three panels: \llep\ 
and \cpd7719 (B1 V).  Bottom three panels: \llep\ and \cpd7724 (B0.5 V). 
All spectra are binned to 0.15 \AA.  Aside from the stronger interstellar 
spectrum and the pronounced high velocity \civ\ absorption in the NGC~6231 
stars, the agreement with \llep\ is very good.}
\label{fig:comp_stds}
\end{center}
\end{figure}

Some years ago, \citet{mas84} noticed that the low dispersion \iue\ spectra 
of several BVs in NGC~6231 had uniquely peculiar spectra, {\em unlike any 
others seen before or since}.  The low resolution \iue\ spectra of these 
stars have abnormally strong \civ$\lambda \lambda 1550$ absorption.  They 
also obtained an \iue\ high dispersion spectrum of the B1 V star, \cpd 7719.  
Later, another high dispersion spectrum of the B0.5 V star, \cpd 7724, was 
also obtained.  The spectra of these stars are compared to the normal 
B0.5~IV star, \llep, in Figure~\ref{fig:comp_stds}.  Because \llep\ was 
used as a flux standard by the \iue\ project, the spectrum shown is a mean 
of 78 spectra.  All of the spectra were binned to 0.15 \AA.  Because \llep\ 
is a slow rotator, $v \sin i = 25$ \kms\ \citep{abt02}, it was necessary to 
broaden its spectrum to obtain good matches to the NGC~6231 stars.  The 
\llep\ spectrum was convolved with a rotational broadening function 
\citep[e.g.,][]{gra76} of 50 \kms\ to match \cpd 7719 and 90 \kms\ to match 
\cpd 7724.  The spectra were also shifted in radial velocity to align the 
Si~{\sc iii}~1300\AA\ triplets.  The agreement of the important Si~{\sc iii} 
1300\AA\ triplets and the 1417\AA\ singlet, which are sensitive to the 
stellar parameters \citep{mas89}, shows that \llep\ is a good spectral match 
for both stars, given the relatively poor S/N of the NGC~6231 spectra.  

The one obvious difference between the NGC~6231 BVs and \llep\ is the high 
velocity \civ\ absorption trough in the NGC~6231 stars, which extends to 
$\sim -2100$ \kms.  In addition, there is no hint of excess emission near 
$v \sim 0$, as one would expect from a normal wind.  These profiles are very 
different from any seen in OB stars of any kind.\footnote{Disk winds 
are also capable of creating absorption with 
little accompanying emission.  However, these winds are typically characterized 
by dense, low speed flows near the star \citep[e.g.,][]{bjo94}, and never show 
high speed absorption without strong low speed absorption.  Thus, it is the 
presence of high speed absorption together with a lack of emission {\em and} 
low speed absorption, that make the profiles of the NGC~6231 BVs unique.} 
\citet{mas84} interpreted 
the peculiar absorption in terms of an abnormal wind.  They were able to 
achieve a good fit to the high resolution profile but, in doing so, it was 
necessary to stretch the bounds of credibility.  
 in order to model .  
Modeling profiles that show very strong high speed blue absorption with 
virtually no red emission with a spherically symmetric wind requires a wind 
that accelerates extremely rapidly, so that nearly all of the wind material 
capable of producing low speed emission is hidden behind the star.  Further, 
this explanation is only plausible if one assumes that carbon is 
overabundant in the BVs.  However, later analysis showed that the NGC~6231 
BVs have fairly normal C abundance \citep[e.g.,][]{kil94}.  

This study included 4 BVs \citet{mas84} identified as peculiar (including 
\cpd 7719), and all were found to have normal C abundances.  In 
addition, over the years since \citet{mas84}, no other examples of BVs with 
similar profiles have emerged \citep[see,][]{wal95}.  As a result, {\em an 
alternative explanation is needed}.  

This paper proposes that the high speed absorption in the BVs actually 
results because they are immersed in a radially expanding flow originating 
elsewhere.  Figure \ref{fig:cluster} shows the stars in the central region 
of NGC 6231.  These are the stars listed in Table 4 of \citet{bau99} and 
\cpd 7719, and includes 10 O stars, a B supergiant and WR~79.  The symbol 
sizes are proportional to the $V$ magnitudes and the coordinate system is 
in arcmin centered on WR~79.  The red points indicate the positions of BVs 
observed in low dispersion by \iue.  This includes the BVs HD~326328, 
HD~326332, HD~326333, \cpd 7715, and Cl*~NGC~6231~SBL, which were observed 
after the data used by Massa et al.\ (1984).  Of these, the spectra of 
\cpd 7715 and  Cl*~NGC~6231 SBL are peculiar by the \citet{mas84} criteria.  
Filled red points are BVs whose low dispersion spectra have peculiar \civ\ 
absorption, and open red points are BVs with normal spectra.  The stars 
labeled 1 and 2 are \cpd 7719 and \cpd 7724, respectively, which have been 
observed in high dispersion by \iue.  It is important to notice that {\em 
the stars with peculiar spectra are those closest to WR~79}.

It is emphasized that such peculiar absorption can only be observed in BVs 
for two reasons.  First, they are strong UV sources at 1540 -- 1550\AA, so 
the absorption can be detected.  Second, unlike more luminous B stars and O 
stars, they do not have strong, high speed absorption from their own winds, 
so it is possible to identify absorption due to a neighboring source.  

\section{Total column through the wind}\label{sec:ew}

This section develops a simple model to help understand why it is so unusual 
to observe the absorption due to a stellar or cluster wind in background 
stars at great distances from the wind source, and why it is possible to see 
such a wind in the case of NGC~6231.  

To begin, consider the absorption caused by a 
spherically symmetric wind when viewed by a second star within the far wind 
(termed the embedded star), which has no wind (or a very weak one), so that 
it can be used as a probe of the strong wind.  Figure~\ref{fig:cartoon} 
shows the coordinate frame where the star with a massive wind is positioned 
at the origin and the embedded star is located at $(x_1, y_1, z_1)$ and the 
observer is at $z = \infty$.  The position of the star is labeled by 
$(p_1, z_1)$, where $p_1$ is the impact parameter, defined as $p = (x^2 
+y^2)^{1/2}$, so that the radial distance between the source and the 
embedded star is $r = (z^2 +p^2)^{1/2}$.  

The region of interest is far 
from the source, where the wind has reached its terminal velocity, 
$v_\infty$.  Therefore, the wind density at $z$ is $\rho(z) = \dot{M}/
[4\pi v_\infty (z^2 +p^2)]$.  The total column density of an absorbing ion 
through the wind to the embedded star is 
\begin{equation}
N_{Tot} = \int_{z_1}^\infty \frac{\rho(z)}{\mu m_H} A_E q_i(z) dz = 
  \frac{\dot{M}A_E q_i}{4\pi \mu m_H v_\infty p} \bigg(\frac{\pi}{2} 
  -\tan^{-1}\frac{z_1}{p}\bigg) \label{eq:ntot}
\end{equation}
where $\mu$ is the mean molecular weight of the gas, $A_E$ is the atomic 
abundance of the absorbing element, $q_i(z)$ is the fraction of the element 
in ionization state $i$ at the location $z$, and $q_i(z)$ was assumed to be 
constant in performing the integration.  Inserting numerical values gives 
\begin{equation}
N_{Tot} = \frac{9.74 \times 10^{15} \dot{M}_{-6}A_E q_i}{v_3 \mu p_{pc}} 
          \bigg(\frac{\pi}{2} -\tan^{-1}\frac{z_1}{p}\bigg)\; {\rm cm}^{-2}
\end{equation}
where $\dot{M}_{-6}$ is the mass loss rate in units of $10^{-6}\; M_\odot$ 
yr$^{-1}$, $v_3$ is the terminal velocity of the gas in thousands of 
km~s$^{-1}$, and $p_{pc}$ is the impact parameter of the line of sight in 
parsecs.  This quantity enters the well known relation for equivalent width, 
$W_\lambda /\lambda_0 = 8.85\times 10^{-21} \lambda_A f N_{Tot}$ 
(e.g., Spitzer 1978), where $\lambda_0$\ is the rest wavelength of the 
transition and $\lambda_A f$ is the same wavelength in \AA\ times the 
oscillator strength of the transition.  

The following analysis concentrates on the C~{\sc iv} $\lambda\lambda 1550$ 
doublet.  For simplicity, this is treated as a single line with a combined 
$\lambda_A f = 442.692$, even though the wavelength ranges of the two 
components do not completely overlap.  With these values the equivalent 
width in \AA\ is
\begin{equation}
W_\lambda = 59.1 \frac{\dot{M}_{-6}A_E q_i}{v_3 \mu p_{pc}}
             \bigg(\frac{\pi}{2} -\tan^{-1}\frac{z_1}{p}\bigg) \label{eq:ew}
\end{equation}
Thus, the absorption seen by an embedded star with $p_{pc} = 1$ and $z_1 = 
0$ (as reference values), due to an O star with a very massive wind 
($\dot{M}_{-6} = 10$, $v_3 \simeq$ 2 -- 3), solar metalicity ($A_E = 3 
\times 10^{-4}$ and $\mu \simeq 1.3$) and the extreme case where all of its 
carbon in \civ\ (i.e., $q_i \simeq 1$), would only be $\sim$ 45 -- 68 m\AA.  
Further, this absorption would be distributed over 10 -- 15 \AA, making it 
impossible to detect.  It would even be extremely difficult to measure 
absorption from a wind 10 times this strong.  However, the situation is very 
different for WR~79.  Because it is a WC star, its wind is carbon rich, with 
$A_E \gtrsim 0.1$ \citep[e.g.,][]{des00}.  As a result, a WC star composed 
of 90\% He and 10\% C ($\mu \sim 4.3$), with $q_i \simeq 1$ and 
$\dot{M}_{-6} \simeq 10$ gives $W_A = 4.6$ -- 6.9 \AA, which is easily 
observed.  Thus {\em the \civ\ equivalent width due to the wind of one WC 
star is stronger than one caused by more than 100 O stars with the same 
\mdot!} 

\section{Model profiles} \label{sec:model}
For a spherically expanding wind, $\rho = \dot{M}/(4 \pi r^2 v)$.  Along a 
sight line with impact parameter $p$, this becomes $\dot{M}/[4 \pi v_\infty 
(p^2 +z^2)]$, where we assume that $p$ is very much larger than the stellar 
radius so that $v = v_\infty$ everywhere along the line of sight.  However, 
what is measured is not $z$, but the line of sight velocity, $v_z$.  From 
Figure~\ref{fig:cartoon}, it is clear that $v_z(z) = v_\infty z/(z^2 
+p^2)^{1/2}$, where $v_z = c (\lambda - \lambda_0)/\lambda_0$, and 
$\lambda_0$ is the rest wavelength of the line in question.  The expression 
for $v_z$ can be inverted to obtain $z(v_z) = v_z p/(v_\infty^2 -v_z^2)^
{1/2}$, and $\frac{dz}{dv_z} = v_\infty^2p/(v_\infty^2 -v_z^2)^{3/2}$.  Now, 
the total column density to a point $z$ in the flow is $\int \rho(z) dz$, 
but this is needed in terms of $v_z$, i.e., $\rho(v_z)_{v_z}$, which is the 
total column density at $v_z$.  Because $\int \rho(z) dz = \int \rho(v_z) 
\frac{dz}{dv_z} dv_z$, one has $\rho(v_z)_{v_z} = \rho(v_z)\frac{dz}{dv_z}$ 
for the density at $v_z$.  Using the expressions for $z(v_z)$ and $\frac{dz}
{dv_z}$ from above, one gets $\rho(v_z) = \dot{M} (v_\infty^2 -v_z^2)/(4 \pi 
v_\infty^3 p^2)$ and 
\begin{equation}
\rho(v_z)_{v_z} = \frac{\dot{M}}{4 \pi v_\infty p (v_\infty^2 
-v_z^2)^{1/2}}
\end{equation}
This is converted to $N(v_z)_{v_z}$, the number of absorbing ions as a 
function of line of sight velocity, 
\begin{equation}
N(v_z)_{v_z} = \frac{9.74\times 10^{15}\dot{M}_{-6} A_E q_i(v_z)}{\mu v_3 
               p_{pc}(v_\infty^2 -v_z^2)^{1/2}} \: \; {\rm cm^{-2} 
               \; {(km \; s^{-1})^{-1}}} 
\end{equation}
for $v_1 \leq v_z \leq v_\infty$, where $v_1$ is the velocity at $z_1$.  
Notice that because the wind extends to $z = \pm \infty$, there 
is always a mathematical singularity at $v = -v_\infty$, and, if $z_1 = 
-\infty$, there will be another at $v = v_\infty$.  
To 
accommodate turbulent velocities along the line of sight, $N(v_z)_{v_z}$ is 
convolved with a normalized Gaussian turbulent velocity profile, $\phi(v_t: 
v_z)$, where $v_t$ is the usual Gaussian $b$ value, i.e., 
$\phi(v_t:v_z) \sim \exp {-[(v -v_z)/v_t]^2}$.
This gives the optical depth as a function of velocity
\begin{eqnarray}
\tau(v_z) & = & \frac{\lambda_0 f}{\frac{m_e c}{\pi e^2}} N(v_z)_{v_z} \star 
            \phi(v_t: v_z) \nonumber \\ 
         & = & \frac{25.8 \lambda_0 f \dot{M}_{-6} A_E}{\mu p_{pc} v_3}
            \left[\frac{q_i(v_z)}{(v_\infty^2 -v_z^2)^{1/2}}
            \star \phi(v_t: v_z)\right] 
\end{eqnarray}

The resulting optical depths of the blue and red components, $\tau_b(v_z)$\ 
and $\tau_r(v_z)$, are then shifted onto a common velocity scale and 
superimposed to produce the total optical depth of the doublet.  Finally, a 
simple change of variables and exponentiation gives the residual intensity 
as a function of wavelength, 
\begin{equation}
r(\lambda) = e^{-[\tau_b(\lambda) +\tau_r(\lambda)]} 
\end{equation}  
which is equivalent to the transmission function of the wind.  

Figure~\ref{fig:profs} shows examples of $r(\lambda)$ for constant 
$q_i(v_z)$.  It demonstrates how the profiles respond to different values 
of $v_1$ and $v_t$.  Note that for the wind of an isolated star, turbulent 
velocities $\sim 0.05$ -- 0.10$v_\infty$ are typically invoked.  Given the 
chaotic nature of the intra-cluster environment and the colliding winds of 
the WR~79 binary, even larger values are not unreasonable. 

\section{Fits to the data} \label{sec:fits}
In this section, the \iue\ high resolution spectra of \cpd 7719 and \cpd 
7724 are fit to the model developed in \S~\ref{sec:model}.  The ionization 
was assumed constant for the fitting procedure, i.e., $q_i(v_z) \equiv q_i$.  
This was necessary since its functional form is unknown and would be 
difficult to disentangle from the velocity parameters which determine the 
shape of the profile with the limited data available.  

Four parameters are used in the fits.  Three are velocity parameters: 
$v_\infty$, the terminal velocity of the absorbing wind; $v_t$, the 
turbulent velocity of the wind, and; $v_1$, the wind velocity at the 
location of the embedded star.  These 3 parameters determine the shape of 
the absorption profile.  The fourth parameter, $u_0$, determines the 
strength of the absorption and is given by 
\begin{equation}
u_0 = \frac{A_E \dot{M}_{-6} q_i}{\mu p_{pc}}
\label{eq:u0}
\end{equation}
which has units of $10^{-6} M_\odot$ yr$^{-1}$ pc$^{-1}$.  

In addition to the absorption profile, a template is needed for the 
unabsorbed flux from the BV.  As shown in Figure~\ref{fig:comp_stds}, the 
B0.5~IV star \llep\ is a good match to both stars, and even the low velocity 
portion of the \civ\ doublets are in reasonable agreement.  Consequently, 
\llep\ was used as a template for both stars.  

Although NGC~6231 has a large number of \bcep\ stars \citep[e.g.,][]{mei13}, 
neither of the program stars have been identified as such.  However, during 
1996, \hst\ GHRS G140L spectra were obtained for \cpd 7719 during 3 
consecutive orbits (spanning roughly 4 hours).  Figure~\ref{fig:ghrs_iue} 
compares the three GHRS spectra with the \iue\ high dispersion spectrum 
degraded to the GHRS G140L spectral resolution of 135 \kms.  Variability is 
clearly present in \civ, and most likely \siiv\ as well.  However, the level 
is much smaller than typically observed in \bcep\ stars.  This may indicate 
that \cpd 7719 is a very low amplitude \bcep, perhaps with such a small 
optical amplitude that it has escaped detection in ground based searches.  
In any case, Figure~\ref{fig:ghrs_iue} demonstrates that exact agreement 
with the low velocity portion of \civ, cannot be expected.  

Figure \ref{fig:fits} shows non-linear least squares fits to \cpd 7719 and 
\cpd 7724 by the absorption model using \llep\ as the underlying 
photosphere.  The fits used the \iue\ errors for the normalized fluxes and 
the points crossed out are contaminated by low velocity interstellar \civ\ 
absorption along the line of sight.  A single $v_\infty$ was used for both 
fits.  For a given fixed value of $v_\infty$, the values of $v_t$, $v_1$ 
and $u_0$ were determined by a Levenberg Marquardt non-linear least squares 
routine.  The best fit value of $v_\infty$ was determined by a simple grid 
search.  In both cases, $\chi^2$ was a minimum for $v_\infty \simeq 1900 \pm 
100$ \kms.  The parameters derived from the fits are listed in 
Table~\ref{tab:params} along with their $1 \sigma$ errors.  Considering the 
data quality data and the simplicity of the model, the fits are considered 
quite good.  

The $v_1$ parameter can be used to infer the $z_1$ distance for each star.  
Assuming a spherical wind from WR~79, and letting $\alpha$ be the angle 
between the line of sight and the flow at $z_1$, one has $v_1 = v_\infty 
\cos \alpha$ and $p_1 = z_1 \tan \alpha$, so $z_1 = p_1 
\cot[\cos^{-1}(v_1/v_\infty)]$.  Using $d = 1.64$~kpc to convert the 
Table~\ref{tab:params} data from arcmin to parsecs, results in $p_1 = 1.3$ 
and $z_1 = 2.20$ pc for \cpd 7719 and 1.1 and 2.23 pc for \cpd 7724.  

Table~\ref{tab:params} also lists the resulting $W_\lambda$ values and 
their errors, which were calculated using the standard propagation of 
errors equation with numerical derivatives for the free parameters of the 
fit.  Using equation~(\ref{eq:ew}), and the values of $z_1$ derived 
above, these can be rescaled to $z_1 = 0$ and $p_{pc} = 1$.  The 
results are 7.66 and 7.61 \AA\ for \cpd 7719 and \cpd 7724, respectively, 
which are close to the expectations given in \S~\ref{sec:ew} for $q_i$ 
\mdot$_{-6} \simeq 10$.

\section{Discussion}\label{sec:dis}
This discussion examines the evidence favoring the proposed interpretation 
of the \civ\ absorption seen in the BVs and describes how the model could 
be used to constrain the kinematics of the flow if more data were available.  
 
To assess the validity of the model, both direct and indirect evidence is 
examined.  Beginning with the direct information, I compare the fit 
parameters ($v_\infty$, $v_t$, $v_1$, $u_0$) given in Table~\ref{tab:params} 
to known quantities and processes relevant to WR stars and H~{\sc ii} 
regions.  The best fit $v_\infty$ (1900 \kms) differs by only 16\% from 2270 
\kms, the $v_\infty$ determined for WR~79 by \citet{pri90}.  The large 
errors for $v_t$ show that they are poorly constrained by the data.  
Nevertheless, values of 300 \kms\ are common for WC stars 
\citep[e.g.,][]{hil89}, so considering the large errors and the chaotic 
nature of the NGC~6231 environment due to the presence of several O stars, 
the derived values for $v_t$ are considered reasonable.  While $v_1$ is not 
directly measurable, it was shown in the previous section that it can be 
used to derive the distance from the star to the wind source.  It was found 
that $z_1 = 2.20$ and 2.23 pc for \cpd 7719 and \cpd 7724, respectively.  
These values are only about twice their impact parameters and, therefore, 
consistent with the observed size of the cluster.  They also indicate that 
the 2 BVs are in front of WR~79.  Combining the $z_1$ and impact parameters 
gives the radial distance of each star from WR~79.  These are 2.57 and 
2.46~pc for \cpd 7719 and \cpd 7724, respectively.  

Knowing that the BVs are 2~pc or more from WR~79, it is possible to estimate 
two important properties about the gas along their lines of sight.  First, 
it takes the absorbing gas $\sim 1000$ yrs to travel from the stellar 
surface to the where it begins to absorb.  Second, The mean number density 
of the wind 2 pc from the star is $0.0032 \;M_{-6}/(\mu r_{pc} v_3) \simeq 
1.57 \times 10^{-4} \;M_{-6}$ cm$^{-3}$. Thus, the absorbing material exists 
in a very rarefied environment. 

The parameter $u_0$, given by equation~(\ref{eq:u0}), is composed of 
quantities that can be constrained by measurements and $q_i$, which cannot.  
Consequently, the process is turned around.  A value for $u_0$ is adopted 
and then it is determined whether reasonable values of $q_i$ result.   To 
begin, equation~(\ref{eq:u0}) is solved for $q_i$  
\begin{equation}
q_i = u_0 p_{pc} \left(\frac{\mu}{A_E}\right) \frac{1}{\dot{M}_{-6}}
\end{equation}
Values for $u_0$ and $p_{pc}$ are set to their means from Table~
\ref{tab:params}, 1.2 pc and 0.31 $10^{-6} M_\odot$ yr$^{-1}$ pc$^{-1}$, 
respectively.  The ratio $\mu/A_E$ depends on the chemical composition or 
the WR~79. If it composed of helium and carbon with $0.08 \leq A_E \leq 
0.25$ \citep{des00}, it will have $4.23 \leq \mu \leq 4.80$ and $0.02 \leq 
A_E/\mu \leq 0.05$.  Previous estimates for \mdot\ of WR~79 are $2.8\times 
10^{-5}$ \mdotperyr\ from its photometric light curve \citep{lam96} and $9 
\times 10^{-5}$ \mdotperyr\ from its radio fluxes \citep{wil91}.  Clearly, 
$q_i$ must be less than one.  The smallest possible value of $q_i$ 
consistent with the data is determined by using the smallest value for 
$\mu/A_E$ and the largest for \mdot$_{-6}$.  This gives $q_i \simeq  0.08$.  
Therefore, for the model to be consistent, $q_i$ must be $\ga 0.08$.  
Unfortunately, wind models offer little information about $q_i$ at such 
large distances from the star, since they rarely extend beyond $\sim 100$ 
stellar radii.  However, most models do predict that the outer wind cools 
to $\sim 10$kK \citep[e.g.,][]{hil89, nug98}, and that nearly all of the C 
is in C~{\sc iii}.  However, the wind material almost certainly contains 
optically thick clumps \citep[e.g.,][and references therein]{ald16}, and at 
such great distances from the star, it must co-exist with the radiation 
field from the cluster O stars and the X-ray emitting gas in the cluster 
\citep{san07}.  These effects are not considered in WR wind models.  As a 
result, $q_i$ for \civ\ is difficult to predict without a better 
understanding of the structure of the flow and detailed modeling, which are 
far beyond the scope of the current paper.  Nevertheless, some guidance can 
be gleaned from interstellar calculations.  First, note that \civ\ is a 
well known tracer of highly ionized gas in H~{\sc ii} regions where, for 
low density environments, $q_i \sim 1$ \citep{dek85}.  Second, \civ\ is 
also an indicator of cooling gas in the temperature range of 80 to 150 kK 
\citep[e.g.,][]{sut93} where $q_i \sim 0.1$.  These conditions are found at 
interfaces between million degree gas and denser clumps 
\citep[e.g.,][]{sav09} -- similar to the conditions in a clumped wind, 
described above.  Consequently, it seems quite plausible that $0.1 \la q_i 
\la 1.0$, depending on exactly how the wind is structured and the influence 
of the surrounding conditions.  

\begin{figure}
\begin{center}
\includegraphics[width=\linewidth]{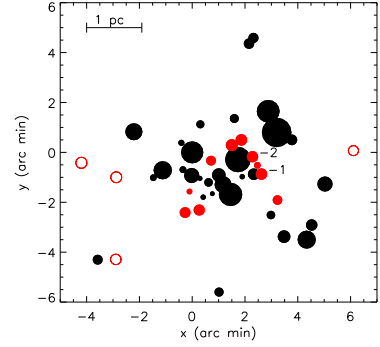}
\vspace{-2.in}\caption{Stars in the central region of NGC 6231.  This 
includes the stars listed in Table 4 of \citet{bau99} and \cpd 7719.  
The size of 1 pc is shown for a distance of 1.64 kpc.  The symbol sizes 
are proportional to $V$ magnitudes and the coordinate system is centered on 
WR~79.  Red points represent BVs observed by \iue.  Filled red points are 
for stars whose low dispersion spectra contain \civ\ peculiarities, and open 
red points are for stars with normal spectra.  The stars labeled 1 and 2 are 
\cpd 7719 and \cpd 7724, respectively, which were observed at high 
resolution with \iue.}
\label{fig:cluster}
\end{center}
\end{figure}

\begin{figure}
\begin{center}
\includegraphics[width=\linewidth]{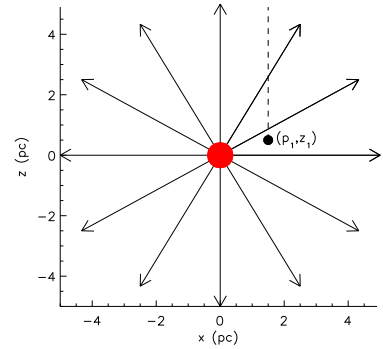}
\end{center}
\vspace{-2.in}\caption{Schematic of a wind emerging from a star at the 
origin and a star embedded in the wind.  The embedded star is located at 
$(x_1, y_1, z_1)$ and labeled by ($p_1, z_1)$, where $p = (x^2 +y^2)^{1/2}$ 
is the impact parameter.  The dashed line points to the observer.  It is 
clear that the range of velocities which absorb along the line of sight is 
dictated by the position of the star in the wind.}
\label{fig:cartoon}
\end{figure}

\begin{figure}
\begin{center}
\includegraphics[width=0.9\linewidth]{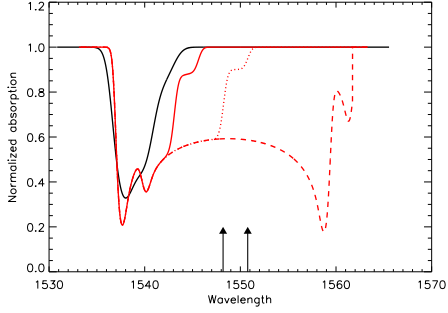}
\end{center}
\vspace{-1.8in}\caption{A set of model profiles of the \civ\ absorption 
expected for embedded stars.  All of profiles have $v_\infty = 2100$ \kms\ 
and the arrows indicate the rest wavelengths of the \civ\ doublet.  The 
black profile has a turbulent velocity of 200 \kms\ and $v_1 = -1500$ \kms, 
implying that the embedded star is well in front of the wind source.  
The red profiles have a turbulent velocity of 100 \kms\ and $v_1 = -1000$ 
\kms (solid), 0 \kms(dotted) and +2100 \kms(dashed).  The profiles with 
$v_1 = 0$ and +2100 \kms\ show how broad, featureless absorption can be 
present and that even red absorption can exist in stars well beyond the wind 
source.}\label{fig:profs}
\end{figure}

\begin{figure}
\begin{center}
\includegraphics[width=\linewidth]{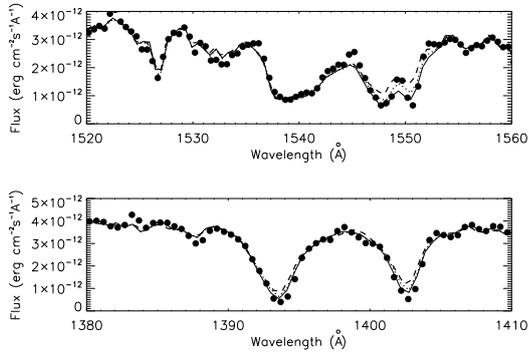}
\vspace{-2.in}\caption{Comparison of 3 GHRS G140L spectra of \cpd7719 
(solid, dashed and dotted curves) taken on consecutive orbits and an \iue\ 
high resolution spectrum binned to the GHRS G140L resolution (points).}
\label{fig:ghrs_iue}
\end{center}
\end{figure}

\begin{figure}
\begin{center}
\includegraphics[width=\linewidth]{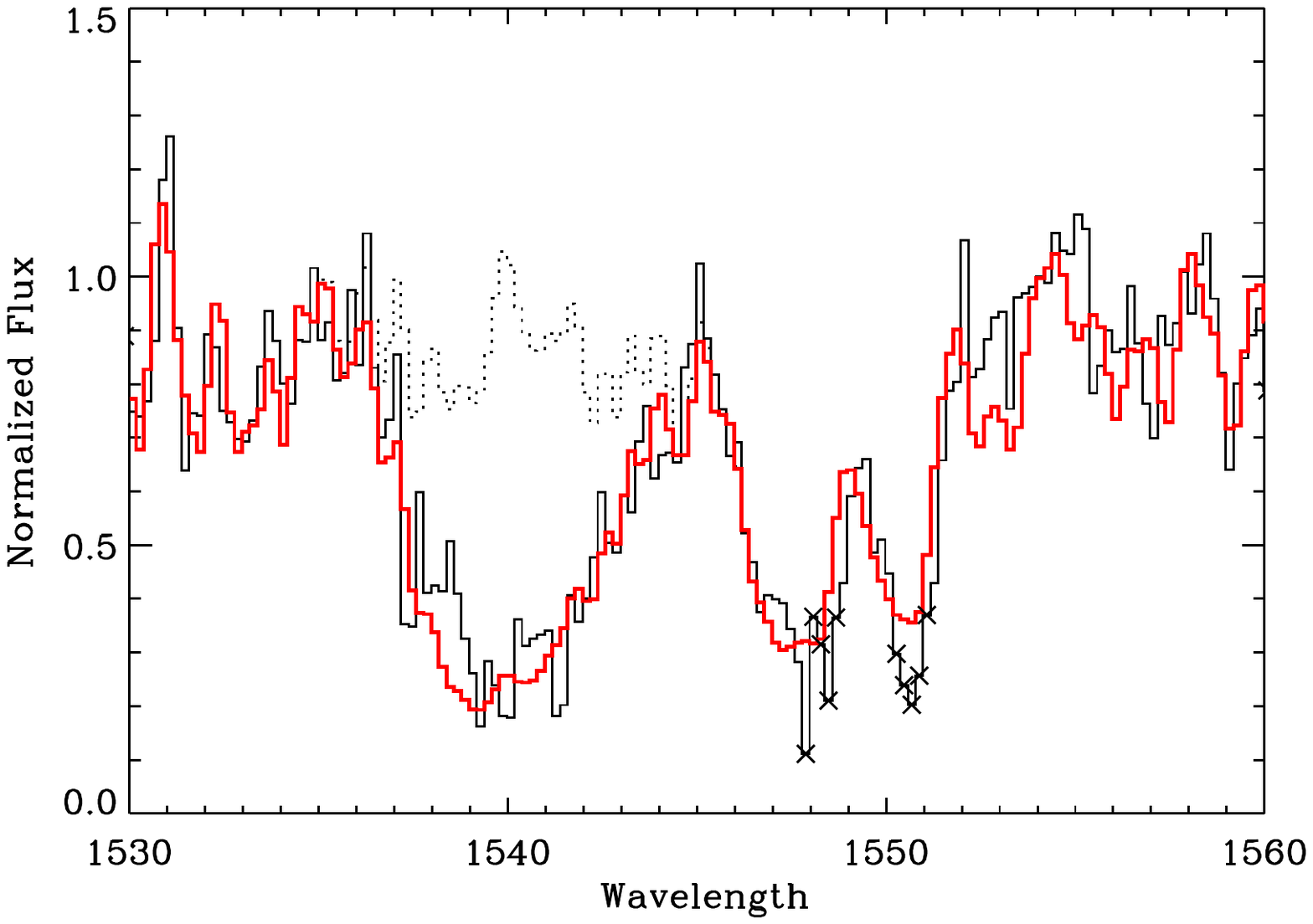}\vfill
\vspace{-2.5in}\includegraphics[width=\linewidth]{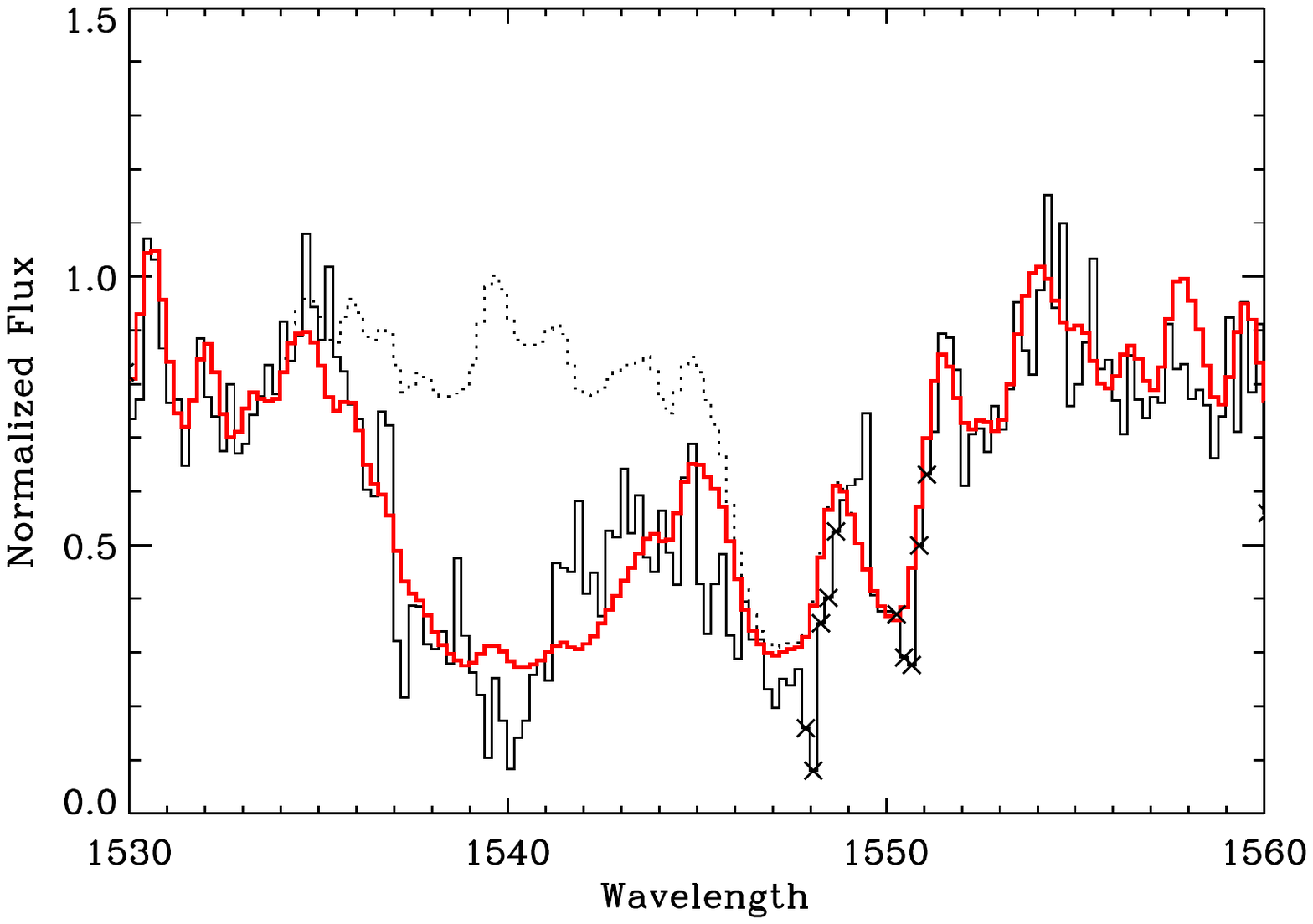}
\vspace{-2.1in}\caption{Fits to \cpd 7719 (left) and \cpd 7724 (right) using 
the parameters listed in Table~\ref{tab:params}. The black spectra are the 
observed spectra, and the crosses represent points excluded from the fits 
due to interstellar contamination.  The dotted spectrum is \llep\ (which was 
used as a template for both stars), and the red spectra are the fits, which 
are the dotted spectrum multiplied by the model absorption profile.}
\label{fig:fits}
\end{center}
\end{figure}

It is also possible that the absorption is not due to just the wind of 
WR~79, but a collective wind whose carbon abundance has been enhanced by 
WR~79.  To address this issue, consider the contributions to the wind flow 
of the 10 O stars (11 when both components of HD 152248 are included) and 
one B supergiant shown in Figure~\ref{fig:cluster}.  Their mass loss rates 
were estimated by beginning with the spectral types given by \citet{bau99}.  
These were translated to physical parameters using the \citet{mar05} 
calibration for Galactic O stars \citep[as modified by][]{wei10} which were 
then used to determine mass loss rates through the \citet{vin00} relations.  
The combined \mdot\ for all of these stars is $1.2\times 10^{-5}$~
\mdotperyr.  While these winds contribute very little to the \civ\ 
absorption, they could influence the wind flow from WR~79.  The two 
components of the binary HD~152248 account for roughly half of the O star 
contribution.  Together, their mass loss is expected to be 6.1$\times 
10^{-6}$ \mdotperyr.  Further, the system is much closer (on the sky) to 
the two BVs (it is the bright star that touches \cpd~7724 in 
Figure~\ref{fig:cluster}).  So although its wind should contribute little 
to the \civ\ absorption, its momentum could have a strong influence on the 
flow that is traced by the \civ\ from WR~79, possibly channeling it toward 
the line of sight to the cluster and affecting the ionization.  
Unfortunately, given the few sight lines available, it is currently not 
possible to pursue this possibility further.  

In terms of indirect support for the current model, consider the statement 
made in \S~\ref{sec:6231} that BVs with anomalous spectra like those in 
NGC~6231 have not been observed before or since in more detail.  
Specifically, beginning with the WC stars listed in the \citet{van01} 
catalog with $V \leq 9$ mag, the \iue\ archive was searched for BVs (with 
an SWP spectrum of any kind) within 30 arcmin of each.  WR~79 is the only 
WC near BVs.  There is a reason for this.  Normal, faint BVs were not 
typically observed with \iue.  Normal BVs are 2 to 4 mags fainter than O, 
B supergiant or WR stars.  Consequently, there was very little motivation 
to observe them, when bright nearby BVs were available.  However, one 
reason to observe faint BVs in open clusters was to use them as standard 
candles for extinction studies -- which is how the BVs with peculiar 
spectra in NGC~6231 were uncovered.  These extinction studies observed 
numerous BVs in several young, open clusters 
\citep[e.g., Fig.~1 in][]{fit07}, but only NGC~6231 contained a WC 
star {\em and} only NGC~6231 contained BVs with peculiar spectra.  Since it 
is now known that the NGC~6231 BVs have normal abundances \citep{kil94}, the 
only thing that makes them unique is that they cohabit a cluster with a WC 
star.  

To summarize, it was argued that the high speed \civ\ absorption is probably 
not intrinsic to the NGC~6231 BVs but related to the wind of WR~79 for the 
following reasons: 
\begin{enumerate}
\item The BVs in NGC~6231 have normal abundances, so there is no reason to 
  expect their winds to be abnormal.  
\item Of the BVs observed in young clusters with \iue, only NGC~6231 
  contains BVs with peculiar \civ\ absorption,  {\em and} only NGC~6231 
  contains a WC star.
\item The NGC~6231 BVs with peculiar \civ\ are located near WR~79, and those 
  further away have normal spectra (\S~\ref{sec:6231}). 
\item The BV absorption profiles are well modeled by the profiles expected 
  for stars embedded in a spherically expanding flow.  
\item The values of $v_\infty$, $v_1$, \mdot$_{-6}$, $A_E$ and $q_i$ 
  determined from the profile fits all lie within the range of expected 
  values. 
\end{enumerate}
While not conclusive, these facts imply that it is highly probable that 
WR~79 is responsible for the high speed absorption.  Unfortunately, the 
available data do not make it possible to determine whether the flow is 
dominated by the wind of WR~79, or simply enriched in carbon by it.  
This is because the two available lines of sight are close together and 
roughly the same distance from WR~79.  As a result, they do not provide 
much independent information for distinguishing whether the high speed gas 
is coming from WR~79 or a more extended source.  Only additional sight 
lines with different locations in the cluster can brake this degeneracy.  

\section*{Acknowledgements}
This paper benefited from the comments of an anonymous referee, and 
discussions with W.-R.\ Hamann and L.\ Oskinova. 
D.M.\ acknowledges partial support under NASA Grants NNX11AB19G and HST 
GO-13760 to SSI.  The data presented in this paper were obtained from the 
Mikulski Archive for Space Telescopes (MAST). STScI is operated by the AURA, 
Inc., under NASA contract NAS5-26555. Support for MAST for non-HST data is 
provided by the NASA Office of Space Science via grant NNX09AF08G and by 
other grants and contracts.  The DSS2 was produced at the STScI, under grant 
NAG W-2166.




\clearpage

\clearpage 

\begin{table}
\begin{center}
\caption{Fit parameters}
\begin{tabular}{llcccccc} \hline
Star    & Sp Ty & $p_1$ & $v_\infty$ & $v_t$ & $v_1$& $u_0$ & $W_\lambda$ \\
        &         & arcmin & \kms   & \kms  & \kms & 
        $10^{-6} M_\odot$ yr$^{-1}$ pc$^{-1}$ &  \AA \\ \hline
\cpd 7719 & B1 V   & 2.8  & $1900 \pm 100$ & $310 \pm 83$ & $1319 \pm 130$ 
                                     & $0.32 \pm 0.05$  & $4.25 \pm 0.57$\\
\cpd 7724 & B0.5 V & 2.3  & $1900 \pm 100$ & $507 \pm 105$ & $1021 \pm 78$ 
                                     & $0.31 \pm 0.05$ & $5.37 \pm 0.58$ \\ 
\hline
\end{tabular}\label{tab:params}
\end{center}
\end{table}


\bsp	
\label{lastpage}
\end{document}